\documentclass[12pt]{article}
\usepackage[a4paper]{geometry}
\usepackage[english]{babel}

\usepackage{amsmath}
\usepackage{amsfonts}
\usepackage{amstext}
\usepackage{amssymb}
\usepackage{amsthm}
\usepackage{amscd}

\usepackage[pagebackref,draft=false]{hyperref}
\hypersetup{colorlinks,
linkcolor=myrefcolor,
citecolor=mycitecolor,
urlcolor=myurlcolor}

\usepackage[capitalize]{cleveref}
\usepackage{caption}
\usepackage{etaremune}

\usepackage{xcolor}
\definecolor{myurlcolor}{rgb}{0,0,0.4}
\definecolor{mycitecolor}{rgb}{0,0.5,0}
\definecolor{myrefcolor}{rgb}{0.5,0,0}
\usepackage{graphicx}
\usepackage{tikz}
\usepackage{tikz-cd}
\usepackage{mathrsfs}

\usepackage{etoolbox}
\usepackage{makeidx}
\usepackage{sectsty}
\usepackage{dsfont}
\usepackage{enumitem} 
\usepackage[]{latexsym}
\usepackage{braket}
\usepackage{caption}
\usepackage[utf8]{inputenx}
\usepackage[T1]{fontenc}
\usepackage{lmodern}
\usepackage{textcomp}
\usepackage{microtype}
\usepackage{totcount}
\usepackage{blindtext}

\newtheorem{remark}{Remark}

\newtheorem*{proof*}{Proof}

\newcommand{\be}{\begin{equation}}
\newcommand{\ee}{\end{equation}}
\newcommand{\bea}{\begin{eqnarray}}
\newcommand{\eea}{\end{eqnarray}}

\title{Descriptions of Relativistic Dynamics with World Line Condition}

\author{F. M. Ciaglia$^{1,6}$  \href{https://orcid.org/0000-0002-8987-1181}{\includegraphics[scale=0.7]{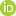}}, F. Di Cosmo$^{2,3,7}$\href{https://orcid.org/0000-0003-0256-5913}{\includegraphics[scale=0.7]{ORCID.png}}, A. Ibort$^{2,3,8}$\href{https://orcid.org/0000-0002-0580-5858}{\includegraphics[scale=0.7]{ORCID.png}}, G. Marmo$^{4,5,9}$\href{https://orcid.org/0000-0003-2662-2193}{\includegraphics[scale=0.7]{ORCID.png}} \\
\footnotesize{$^{1}$\textit{ Max Planck Institute for Mathematics in the Sciences, Leipzig, Germany}} \\
\footnotesize{$^{2}$\textit{ ICMAT, Instituto de Ciencias Matem\'{a}ticas (CSIC-UAM-UC3M-UCM)}}  \\
\footnotesize{$^{3}$\textit{ Depto. de Matem\'aticas, Univ. Carlos III de Madrid, Legan\'es, Madrid, Spain}}  \\
\footnotesize{$^{4}$\textit{ Dipartimento di Fisica ``E. Pancini'', Universit\`a di Napoli Federico II, Napoli, Italy}} \\
\footnotesize{$^{5}$\textit{ INFN-Sezione di Napoli, Napoli, Italy.}} \\
\footnotesize{$^{6}$\textit{ e-mail: \texttt{florio.m.ciaglia[at]gmail.com}}, $^{7}$\textit{ e-mail: \texttt{fcosmo[at]math.uc3m.es}}, } \\
\footnotesize{$^{8}$\textit{ e-mail: \texttt{albertoi[at]math.uc3m.es}}, $^{9}$\textit{ e-mail: \texttt{marmo[at]na.infn.it}}}
}

\date{}

\begin{document}

\maketitle

\abstract{In this paper a generalized form of relativistic dynamics is presented: A realization of the Poincaré algebra is provided in terms of vector fields on the tangent bundle of a simultaneity surface in $\mathbb{R}^4$. The construction of this realization is explicitly shown to clarify the role of the commutation relations of the Poincaré algebra versus their description in terms of Poisson brackets in the no-interaction theorem. Moreover, a geometrical analysis of the ``eleventh generator'' formalism introduced by Sudarshan and Mukunda is outlined, this formalism being at the basis of many proposals which evaded the no-interaction theorem.}

\vspace{0.3cm}

\textit{In memory of E.C.G. Sudarshan, who was interested, for almost three decades, in problems of covariant description of relativistic interacting particles.}

\section{Introduction}
\label{sec-Introduction}
Seventy years ago, Dirac \cite{dirac-forms_of_relativistic_dynamics} argued that a covariant description of relativistic dynamics means to find a realization of the Lie algebra of the Poincaré group in terms of observables and a Poisson bracket. In the so-called instant form, Dirac argued that the role of the single Hamiltonian in non relativistic dynamics should be replaced by four Hamiltonians, one of them being the total energy of the system, the other three being the boosts. Quoting Dirac: «\textit{The equations of motion should be expressible in the Hamiltonian form. This is necessary for a transition to the quantum theory to be possible}».

All subsequent papers, therefore, always assumed the realization of the Poincaré algebra to be given in terms of Poisson Brackets on dynamical variables on a carrier space which would attribute three positions and three momenta (or velocities) to each particle. The additional requirement that every particle evolution would be associated with an invariant world-line on space-time, gave rise to the world-line condition and, in the hands of Sudarshan and collaborators, ended up with the no-interaction theorem \cite{currie_jordan_sudarshan-relativistic_invariance_and_hamiltonian_theories_of_interacting_particles}. For the sake of clarity, let us recall what are the assumptions at the base of this theorem: The physical system is described in the canonical formalism by a phase space with an associated Poisson bracket; the three-dimensional position coordinates of each particle at a common physical time (instant form of dynamics) represent half of a system of canonical variables for this phase space; under the Euclidean subgroup of the Poincaré group (characteristic of the instant form) the canonical and geometrical transformation laws for these coordinates coincide; if in any state of motion, as seen from a given reference frame, the world-lines of the particles are drawn in space-time, then the canonical transformations which relates this description with the one seen from a different frame preserve the objective reality of these lines. These conditions express the two aspects of relativistic invariance in the description of a physical system: From one point of view physical laws have to be invariant under changes of reference frames (``relativistic invariance''); on the other hand some physical quantities transform in a specific way under changes of reference frames due to the action of the Lorentz group on space-time events (``manifest invariance''). Then the no-interaction theorem states that, when considering a system of particles, the only dynamical evolution which is compatible with the above assumptions is the free dynamical evolution. This conclusion is in agreement with the relativistic principle of constancy of the speed of light, which forbids instantaneous interactions propagating faster than the speed of light. These results have supported the development of field theories: Particles locally exchange energy and momenta with fields which have their own degrees of freedom and their own dynamics (For a detailed discussion about the meaning of an action-at-a-distance in the context of classical and quantum mechanics we refer to \cite{sudarshan-action_at_a_distance}, where a comparison with local field theories is also presented). 

Later on, in the eighties, many relativists and particle physiscists took up the problem of providing a covariant description of relativistic interacting particles without the intervention of fields, i.e., a kind of ``action-at-a-distance'' covariant under the Poincaré group.In these attempts (see, for instance \cite{komar-interacting_relativistic_particles,todorov-dynamics_of_relativistic_point_particles_as_a_problem_with_constraints,rohrlich-relativistic_hamiltonian_dynamics_I}) interactions were described via constraints, using Dirac description of Hamiltonian constrained systems \cite{dirac-lectures_on_quantum_mechanics}. In the beginning these models were supposed to violate the world line condition (WLC). However, Sudarshan and Mukunda proposed an interpretation in terms of an ``eleventh generator'' formalism \cite{mukunda_sudarshan-forms_of_relativistic_dynamics_with_world_lines}. The set of constraints used to define the interaction among particles depend on an additional variable, parameterising the curves which constitute a state of the motion for the system. The generator of this ``evolution'' is independent of the other ten generators of the Poincaré group, obtaining the so-called ``eleventh generator'' formalism. They also noticed that the notion of WLC can be meaningfully implemented in this framework, providing examples of interacting particles satisfying the WLC. The difference with the assumptions of the no-interaction theorem, indeed, consists in the ``dynamical'' choice of the time parameter: In the constrained formalism the time depends on the state of the motion of the system and there is not a neat separation between kinematics and dynamics, as in Dirac's form of relativistic dynamics. In other words, the variables interpreted as the world line positions after implementing all the constraints do not need to coincide with the canonical positions of the phase space initially associated with the physical system. Later on \cite{balachandran_marmo_mukunda_nilsson_simoni_sudarshan_zaccaria-unified_geometrical_approach_to_relativistic_particle_dynamics} it was proven that all these constrained descriptions can be derived derive from a common ``covering phase space'' via suitable reduction procedures. 

However, the additional requirement of separability, i.e., the requirement that in the instant form clusters of particles very far apart should behave as non-interacting, gave rise to a novel version of the no-interaction theorem \cite{balachandran_dominici_marmo_mukunda_nilsson_samuel_sudarshan_zaccaria-separability_in_relativistic_hamiltonian_particle_dynamics}. All these results pointed out again that a suitable covariant description of relativistic interacting particles cannot be formulated without the intervention of fields. 

With the hope to clarify the role of the commutation relations of the Poincaré algebra versus their description in terms of Poisson Brackets, in this paper we propose a solution of the problem posed by Dirac in terms of vector fields, we call it a Newtonian realization, the dynamical vector field being a second order one, and then we require this realization to be compatible with a Lagrangian description (a covariant inverse problem for the full Poincaré algebra, not just the dynamcs). We also provide a geometrical description of the ``eleventh generator'' formalism by means of Jacobi Brackets. We find that in this approach ``canonical positions'' do not coincide with ``geometrical positions'', i.e., positions in space-time. We shall present these various aspects in the particular case of one particle relativistic systems with possible ``external forces'', a situation which would arise after implementing the ``separability condition''. 

\section{A geometrical formulation of Dirac's problem} \label{sec.2}

In modern geometrical terms, the problem formulated by Dirac may be formulated as follows. The Lie agebra of any Lie group $G$, say $l_G$, canonically defines a Poisson structure on the dual space $l^*_G = Lin(l_G, \mathbb{R})$ by means of the following construction. With every element $u \in l_G$, we associate a linear function $\hat{u} \in \mathcal{F}(l_G^*)$, and we define a Poisson Bracket on $\mathcal{F}(l_G^*)$ by setting
\begin{equation}
\left\lbrace \hat{u}, \hat{v} \right\rbrace = \widehat{\left[ u , v \right]}\,,
\end{equation} 
or, if $\alpha \in l_G^*$
\begin{equation}
\left\lbrace \hat{u}, \hat{v} \right\rbrace (\alpha) = \alpha( \left[ u , v \right])\,.
\end{equation} 
This bracket leads to a well defined tensor field even in infinite dimensions provided that $l_G$ may be identified with its double dual $(l_G^*)^*$, that is, whenever  $l_G$ is reflexive. Thus, if $P$ denotes the Poincaré group, we introduce the following notations for the elements of $l_{P}$:
\begin{equation}\bibliographystyle{plain}
P_{\mu}\,, \qquad \mu = 0,1,2,3
\end{equation} 
form the Abelian part of the Poincaré algebra, and
\begin{equation}
M_{\mu \nu}\,, \qquad \mu,\,\nu = 0,1,2,3\,, 
\end{equation}  
represents rotations and boosts. Then, the commutation relations read
\begin{eqnarray}
& \left[ P_{\mu} , P_{\nu} \right]= 0 \\
& \left[ M_{\mu \nu} , P_{\rho} \right] = g_{\nu \rho}P_{\mu} - g_{\mu \rho}P_{\nu}\\
& \left[ M_{\mu \nu} , M_{\rho \sigma} \right] = g_{\nu \rho}M_{\mu \sigma} - g_{\mu \rho}M_{\nu \sigma} + g_{\nu \sigma}M_{\rho \mu} - g_{\mu \sigma}M_{\rho \nu}\,,
\end{eqnarray}
where $g_{\mu \nu}$ is the Minkowski metric on $\mathbb{R}^4$, and they can be used to explicitly characterize also the Poisson algebra on the dual space $l^*_P$. 

Now, the problem as stated by Dirac amounts to find a ``symplectic realization'', or a ``Lagrangian realization'' \cite{abraham_marsden-foundations_of_mechanics,giordano_marmo_simoni-symplectic_and_lagrangian_realization_of_poisson_manifolds} of the Poisson manifold $\left( l^*_P, \left\lbrace \cdot , \cdot \right\rbrace \right)$, associated with the Poincaré group. This means we need to find a Poisson map:
\begin{equation}
\mu \, : \, T^*Q \, \rightarrow \, l^*_P
\end{equation} 
in the phase-space case, or a Poisson map
\begin{equation}
\mu_{\mathcal{L}} \, : \, TQ \, \rightarrow \, l^*_P
\end{equation} 
in the Lagrangian one. In the latter situation, we have to consider a Lagrangian-dependent Poisson bracket, since the tangent bundle is not equipped with a canonical symplectic structure.  
\begin{remark}
This formulation of the problem would be consistent with Wigner's approach to the classification of elementary particles as irreducible unitary representations of the Poincaré group. The classical counterpart would correspond to the coadjoint orbits of the Poincaré group acting on the dual of the Lie algebra.
\end{remark}

Our ``Newtonian realization'' would correspond, instead, to a map:
\begin{equation}
\nu \, : \, l_P \, \rightarrow \, \chi(\mathcal{M})\, 
\end{equation} 
such that the Lie bracket in the Lie algebra of the Poincaré group is realized in terms of the commutator between the vector fields on some carrier space $\mathcal{M}$.
Note that a similar idea has been developed in \cite{grig1992}. 

\section{Newtonian realization in the instant form}\label{sec.3}
Let us start with the vector space $\mathbb{R}^4$ with globally defined coordinate functions $x^{\mu}$, $\mu = 0,1,2,3$, and  equipped with the Minkowski metric tensor 
$$g = -dx^0\otimes dx^0 + \sum_{j=1}^3dx^j\otimes dx^j.$$
In this space, a simultaneity surface for an inertial reference frame is diffeomorphic with $\mathbb{R}^3$, that is, it represents the space of positions at a given instant. The coordinate functions for this copy of $\mathbb{R}^3$ are given by the $x_j$'s with $j=1,2,3$.  Positions and velocities are points of the tangent bundle $T\mathbb{R}^3$, which will be our carrier space $\mathcal{M}$. Thus, we have to find a realization of the Poincaré algebra in terms of vector fields on $T\mathbb{R}^3$. 

It is clear that, since space rotations and space translations form the Euclidean group which preseves every simultaneity leaf associated with a given inertial frame, we can exploit the tangent lift of their standard realization on $\mathbb{R}^{3}$. On the other hand, in Dirac parlance, the complicated generators will be those corresponding to the ``Hamiltonians'', i.e., the dynamics and the three boosts. It is at this point that the world-line condition plays a preminent role. Let $\Gamma$ denote the second order dynamical vector field, and $K_1, K_2, K_3$ the vector fields representing the boosts. Then, we require that the following conditions hold:
\begin{eqnarray}
& L_{K_j}x_l = x_j L_{\Gamma}x_l = x_j \dot{x}_l \\
& L_{K_j}\dot{x}_l = \dot{x}_j\dot{x}_l + x_j L_{\Gamma}\dot{x}_l - \delta_{jl}\,. 
\end{eqnarray} 
These vectors fields express the world-line condition as stated by Sudarshan and collaborators\cite{marmo_mukunda_sudarshan-relativistic_particle_dynamics_lagrangian_proof_of_the_no_interaction_theorem}.

Moreover, we notice that $K_j$ is a non-linear vector field, it does not respect the tangent bundle structure of $T\mathbb{R}^3$, and it is the sum of a Newtonoid \cite{marmo_ferrario_lovecchio_morandi_rubano-the_inverse_problem_in_the_calculus_of_variations_and_the_geometry_of_the_tangent_bundle} vector field and a vertical lift of a translation for all $j=1,2,3$. Indeed, we have
\begin{equation}
K_j = x_j \dot{x}_l\frac{\partial}{\partial x_l} + \dot{x}_j \dot{x}_l\frac{\partial}{\partial \dot{x}_l} + x_j L_{\Gamma}\dot{x}_l \frac{\partial}{\partial \dot{x}_l} - \delta_{jl}\frac{\partial}{\partial \dot{x}_l} = x_j \Gamma + \dot{x}_j \Delta - \delta_{jl}\left( \frac{\partial}{\partial x_l} \right)^V\,,
\end{equation}
where $\Delta = \dot{x}_j\frac{\partial}{\partial \dot{x}_j}$ is the dilation vector field along the fibers of the tangent bundle $T\mathbb{R}^3$ \cite{carinena_ibort_marmo_morandi-geometry_from_dynamics_classical_and_quantum,marmo_ferrario_lovecchio_morandi_rubano-the_inverse_problem_in_the_calculus_of_variations_and_the_geometry_of_the_tangent_bundle} and $\left(\frac{\partial}{\partial x_j} \right)^V$ denotes the vertical lift of the vector field $\frac{\partial}{\partial x_j}$ \cite{kolar_slovak_michor-natural_operations_in_differential_geometry}. In summary, the Newtonian realization of the Poincaré algebra is given by 
\begin{eqnarray}
& P_j = -\frac{\partial }{\partial x_j}\,, \qquad P_0 = \Gamma\\
& J_l = \epsilon_{ljk}\left( x_j \frac{\partial}{\partial x_k} + \dot{x}_j \frac{\partial}{\partial \dot{x}_k} \right)\,, \quad K_j = x_j \Gamma + \dot{x}_j \Delta + \left( P_j \right)^V\,.
\end{eqnarray}
We eventually notice that this realization of the Poincaré algebra is a dynamical one, i.e., it depends on the second order dynamics $\Gamma$.

By requiring these vector fields to satisfy the commutation relations of the Poincaré algebra, we get a system of partial differential equations for the accelerations, i.e., we derive a system of PDE for the functions $a_1,a_2,a_3$ which appear in the second order dynamical vector field $$\Gamma = \dot{x}_j \frac{\partial }{\partial x_j} + a_j \frac{\partial }{\partial \dot{x}_j}\,.$$

In particular, the commutation relations with the spatial translations impose that the accelerations cannot depend on the positions $x_j$. The commutation relations with the rotations, instead, imply the following expression:
\begin{equation}
a_j = \dot{x}_j f(\dot{x}^2)\,.
\end{equation}
Finally, let us consider the commutation relation $[K_1,K_2]=J_3$. This relation is verified if and only if the following set of PDE holds true:
\begin{eqnarray}
&2(x_1\dot{x}_2 - x_2\dot{x}_1)\dot{x}_3\left[ (1-\dot{x}^2)\frac{\partial f}{\partial \dot{x}^2} + f\right] = 0  \\
& -x_2 f +2 (x_1\dot{x}_2 - x_2\dot{x}_1)\dot{x}_1\left[ (1-\dot{x}^2)\frac{\partial f}{\partial \dot{x}^2} + f\right] = 0 \\
& x_1 f +2 (x_1\dot{x}_2 - x_2\dot{x}_1)\dot{x}_2\left[ (1-\dot{x}^2)\frac{\partial f}{\partial \dot{x}^2} + f\right] = 0\,.
\end{eqnarray}
The only solution to this system is $f=0$, which means that the only dynamics, as intended by Dirac, compatible with the world-line condition must admit a generator $\Gamma$ which has vanishing accelerations. 

An additional result can be obtained if one requires that the realization by  means of vector fields allows for a compatible non-trivial Lagrangian function \cite{balachandran_marmo_stern-a_lagrangian_approach_to_the_no_interaction_theorem,marmo_mukunda_sudarshan-relativistic_particle_dynamics_lagrangian_proof_of_the_no_interaction_theorem}, i.e., for any of the vector fields in the Newtonian realization the following condition must be satisfied:
\begin{equation}
L_X\omega_{\mathcal{L}} = 0\,,
\end{equation} 
or, equivalently, 
\begin{equation}
L_X\theta_{\mathcal{L}} = dF_X\,,
\end{equation} 
due to the contractability of $T\mathbb{R}^3$. By using $L_{K_m}\theta_{\mathcal{L}} = dF_m$, $L_{\Gamma}\theta_{\mathcal{L}} = d\mathcal{L}$, $L_{P_m}\theta_{\mathcal{L}} = L_{J_m}\theta_{\mathcal{L}} = 0$ we find that
\begin{equation}
L_{K_m}\mathcal{L} = L_{\Gamma}F_m\,, \quad m=1,2,3\,.
\label{generating function boost}
\end{equation}
Since $\Gamma$ is the vector field describing the dynamic of a non-interacting particle, we can find a solution for $F_m$ having the form
\begin{equation}
F_m = x_m h(\dot{x}^2)\,,
\end{equation} 
where $\dot{x}^2 = \sum_j \dot{x}_j \dot{x}_j$ and we can add to it any constant of the motion. Then Eq.\eqref{generating function boost} becomes
\begin{equation}
2\left( \dot{x}^2 -1 \right) \frac{\partial \mathcal{L}}{\partial \dot{x}^2} = h\left( \dot{x}^2 \right)\,.
\end{equation} 
The commutation relations of the Poincaré algebra, however, impose some additional conditions. Indeed, since 
\begin{equation}
\left[ K_m , P_n \right] = \delta_{mn} \Gamma\,,
\end{equation}
we have that 
\begin{equation}
L_{P_m}L_{K_n}\theta_{\mathcal{L}} = -\delta_{mn}L_{\Gamma}\theta_{\mathcal{L}}\,,
\end{equation}
which implies the following equation for $F_n$
\begin{equation}
L_{P_m}F_m = -\mathcal{L}\,.
\end{equation}
This is an equation involving only the Lagrangian function 
\begin{equation}
2(\dot{x}^2 - 1)\frac{\partial \mathcal{L}}{\partial \dot{x}^2} = \mathcal{L}\,,
\end{equation}
which integrates to
\begin{equation}
\mathcal{L} = c\sqrt{1-\sum_j (\dot{x}_j)^2}\,,
\end{equation}
with $c\in \mathbb{R}$. 


This result is quite remarkable because, by requiring the Newtonian realization of the Poincaré algebra to allow for a Lagrangian description, we obtain  that the Lagrangian is unique. This uniqueness property is very startling when we recall that, if we only require the dynamical vector field 
\begin{equation}
\Gamma = \dot{x}_j\frac{\partial}{\partial x_j}
\end{equation}
to allow for a Lagrangian description, we find an infinite family of solutions provided not only by any function $\mathcal{L} = \mathcal{L}(\dot{x}^2)$, but also $\mathcal{L} = \mathcal{L}(\dot{x}_1,\,\dot{x}_2,\,\dot{x}_3)$.
In other words, among the infinitely many Lagrangian functions providing a description of the free dynamics, there is only one which admits the above Newtonian realization of the Poincaré algebra as generalized Noether symmetries.

\section{The eleventh-generator formalism} \label{sec.4}
In \cite{mukunda_sudarshan-forms_of_relativistic_dynamics_with_world_lines}, Mukunda and Sudarshan introduced an eleventh-generator formalism. We would like to unveil the geometric content of their formalism because it forms the prototype for the proposals made in the eighties to evade the no-interaction theorem by means of the Dirac-Bergmann constraint formalism \cite{komar-interacting_relativistic_particles,rohrlich-relativistic_hamiltonian_dynamics_I,todorov-dynamics_of_relativistic_point_particles_as_a_problem_with_constraints}. 

We start with what Dirac calls an elementary solution of the symplectic realization of the Poisson manifold defined by $l_P^*$, that is, the phase-space $T^*\mathbb{R}^4$ with coordinates $(x_{\mu}, p_{\mu})$, and with symplectic structure given by the natural one, 
\begin{equation}
\omega = dp_{\mu}\wedge dx^{\mu} = g^{\mu \nu} dp_{\mu}\wedge dx_{\nu}\,,
\end{equation} 
which is the exterior differential of the one form 
\begin{equation}
\theta_0 = p_{\mu}dx^{\mu} = g^{\mu \nu}p_{\mu}dx_{\nu}\,.
\end{equation} 
By setting $P_{\mu}  = p_{\mu}$, $M_{\mu \nu}=x_{\mu}p_{\nu} - x_{\nu}p_{\mu}$ we obtain Dirac's elementary solution of the problem in terms of Poisson Brackets and generating functions.

Starting from this solution, we can construct another solution by adding an element to the Poincaré algebra. Specifically, on $T^*\mathbb{R}^4$ we consider the submanifold 
\begin{equation}
\Sigma_m = \left\lbrace (x_{\mu}, p_{\mu})| p_{\mu}p^{\mu}=m^2 \right\rbrace\,.
\end{equation}
If we consider the natural immersion of $\Sigma_m$ into $T^*\mathbb{R}^4$, it is possible to write the pull-back 
\begin{equation}
i^*_{\Sigma_m}\theta_0 = \theta_m\,. 
\end{equation}
On $\Sigma_m$, the form $\theta_m$ defines a contact structure \cite{arnold-mathematical_methods_of_classical_mechanics,asorey_ciaglia_dicosmo_ibort_marmo-covariant_jacobi_brackets_for_test_particles}, since $\theta_m \wedge (d\theta_m)^3 \neq 0$ represents a volume form. Then, it is possible to define a Lie algebra structure \cite{kirillov-local_lie_algebras} on $\mathcal{F}(\Sigma_m)$ by setting
\begin{equation}
\left[ f , g \right]_m \theta_m \wedge (d\theta_m)^3 = (fdg - gdf)\wedge (d\theta_m)^3 + 2 df\wedge dg \wedge \theta_m \wedge (d\theta_m)^2\,. 
\end{equation}
The given expression shows that the bracket, being defined only in terms of $\theta_m$ which is Poincaré invariant, is indeed preserved by the action of the Poincaré group given above. As a matter of fact, this bracket can be described in the more common setting of a bivector field and a vector field, say $(\Lambda_m, \Gamma_m)$, defined by 
\begin{eqnarray}
& i_{\Gamma_m} \left( \theta_m \wedge (d\theta_m)^3 \right) = \left( d\theta_m \right)^3 \\
& i_{\Lambda_m} \left( \theta_m \wedge (d\theta_m)^3 \right) = 3 \theta_m \wedge \left( d\theta_m \right)^2\,.
\end{eqnarray}
They allow the definition of the previous bracket as follows
\begin{equation}
\left[ f,g \right]_m = \Lambda_m(df,dg) + fL_{\Gamma_m}g - gL_{\Gamma_m}f\,,  
\end{equation}
and the Jacobi identity holds true because of the following properties:
\begin{eqnarray}
& \left[ \Lambda_m , \Lambda_m \right] = 2 \Gamma_m \wedge \Lambda_m \\
& \left[ \Gamma_m , \Lambda_m \right] = 0\,, 
\end{eqnarray}
where the bracket between multivector fields is the Schouten bracket \cite{kolar_slovak_michor-natural_operations_in_differential_geometry}.

By using the pair $\left( \Lambda_m, \Gamma_m \right)$, it is possible to associate with any function $f$ a first order differential operator $\tilde{X}_f$ and a vector field $X_f$, respectively given by
\begin{eqnarray}
& \tilde{X}_f = \Lambda_m(df, \cdot) + f\Gamma_m - L_{\Gamma_m}f \\
& X_f = \Lambda_m(df,\cdot) + f\Gamma_m\,.
\end{eqnarray}
It should be noticed that constant functions are not mapped into the null vector field, but in both cases we have $X_c = c\Gamma_m$, for any $c\in \mathbb{R}$. Moreover, the Lie bracket does not satisfy the Leibniz rule, but we have  
\begin{equation}
\left[ f, gh \right]_m = \left[ f, g \right]_m h + g \left[ f, h \right]_m - \left[ f, 1 \right]_m gh\,,
\end{equation}
showing an important difference between $\left[ f, g \right]_m$ and $\left\lbrace f,g \right\rbrace$, where the first is called Jacobi bracket and the second Poisson Bracket. It is common to call $X_f$ the Hamiltonian vector field associated with $f$. However, on the subalgebra of functions which are constants of the motion for $\Gamma_m$, i.e., $L_{\Gamma_m}f=0$, the Jacobi bracket reduces to a Poisson Bracket. 

It turns out that the usual generating functions of the Poincaré algebra given in $T^*\mathbb{R}^4$, when pulled back to $\Sigma_m$, provide a solution of the Dirac problem because they are constants of the motion for $\Gamma_m$ and therefore generate a realization of the Poincaré algebra in terms of Poisson Brackets. To compare the Jacobi algebra with the Poisson Bracket on $T^*\mathbb{R}^4$, we consider the following bivector field 
\begin{equation}
\Lambda = \left( g^{\mu \nu} - \frac{p^{\mu}p^{\nu}}{p_{\mu}p^{\mu}} \right) \frac{\partial}{\partial p^{\mu}} \wedge \frac{\partial}{\partial x^{\mu}} = g^{\mu \nu} \frac{\partial}{\partial p^{\mu}} \wedge \frac{\partial}{\partial x^{\mu}} - \Delta \wedge \Gamma\,,
\end{equation}  
where
\begin{equation}
\Delta = p^{\mu}\frac{\partial}{\partial p^{\mu}}\,,\quad \Gamma= \frac{p^{\mu}}{p_{\nu}p^{\nu}}\frac{\partial}{\partial x^{\mu}}\,.
\end{equation}
All these contravariant tensor fields are actually tangent to the leaves of the foliation defined by $p_{\mu}p^{\mu}=m^2$ when $m$ changes and we remove the manifold defined by $p_{\mu}p^{\mu}=0$. Then, on each mass-shell $\Sigma_m$, we have the following pair
\begin{eqnarray}
& \Lambda_m = \left( g^{\mu \nu} - \frac{p^{\mu}p^{\nu}}{m^2} \right) \frac{\partial}{\partial p^{\mu}} \wedge \frac{\partial}{\partial x^{\nu}} \\
& \Gamma_m = \frac{p^{\mu}}{m^2}\frac{\partial}{\partial x^{\mu}}\,,
\end{eqnarray} 
and the associated Jacobi bracket is given by the following relations:
\begin{eqnarray}
&\left[ x^{\rho} , x^{\sigma} \right]_m = \frac{x^{\rho}p^{\sigma} - x^{\sigma}p^{\rho} }{m^2} \\
& \left[ p^{\rho} , x^{\sigma} \right]_m = g^{\rho \sigma} \\
& \left[ p^{\rho} , p^{\sigma} \right]_m = 0\,.
\end{eqnarray} 
By using the generating functions which are constants of the motion for $\Gamma$, i.e., $M_{\mu \nu}$ and $P_{\mu}$, we find the associated vector fields:
\begin{eqnarray}
& X_{\mu \nu} = x_{\mu}\frac{\partial }{\partial x^{\nu}} - x_{\nu}\frac{\partial }{\partial x^{\mu}} + p_{\mu}\frac{\partial }{\partial p^{\nu}} - p_{\nu}\frac{\partial }{\partial p^{\mu}} \\
& Y_{\mu} = \frac{\partial }{\partial x^{\mu}}\,. \\ 
\end{eqnarray}
Therefore, in this realization, the Poincaré algebra, represented in terms of vector fields, contains a central element given by $\Gamma_m$ which plays the role of the eleventh generator.

\section{A ``frozen phase-space'' realization} \label{sec.5}

In the same spirit of the previous section, we start with the phase-space $T^*\mathbb{R}^4$, we remove the manifold defined by $p_{\mu}p^{\mu}=0$, and we consider the one-form 
\begin{equation}
\theta = \frac{\theta_0}{\sqrt{p_{\mu}p^{\mu}}}\,,
\end{equation}
which, on a given mass-shell $\Sigma_m$, would be
\begin{equation}
\tilde{\theta}_m = \frac{\theta_m}{m}\,.
\end{equation}

Then, the two-form $d\theta = \frac{d\theta_0}{\sqrt{p_{\mu}p^{\mu}}} -\frac{p_{\mu}dp^{\mu}}{\sqrt{p_{\mu}p^{\mu}}}\wedge \theta_0 $ is a degenerate two-form whose kernel is generated by
\begin{equation}
\Delta = p^{\mu}\frac{\partial}{\partial p^{\mu}}\quad \mathrm{and} \quad \Gamma= \frac{p^{\mu}}{p_{\nu}p^{\nu}}\frac{\partial}{\partial x^{\mu}}\,.
\end{equation} 
Indeed, $\theta$ is invariant under dilation because it is homogeneous of degree zero in the momenta. The fact that $\Gamma$ is in the kernel of $d\theta$ follows by direct computation. 

As the infinitesimal generators which realize the Poincaré algebra commute with $\Delta$ and $\Gamma$, they descend to the quotient manifold which is symplectic and six dimensional. We obtain in this way another solution of Dirac's problem in terms of Poisson brackets. 
However, there is no evolution on points of this quotient manifold because we quotiented out the dynamics represented by $\Gamma$, and therefore, this realization is on a ``frozen phase-space'' as it was called by P.Bergmann and A.Komar \cite{komar_bergmann-frozen}. 

\section{A Lagrangian solution to the Dirac problem}

In this last section, it is useful to show how it is possible to provide a realization of the Poincaré algebra in the Lagrangian formalism on $T\mathbb{R}^4$ \cite{dubrovin_giordano_marmo_simoni-poisson_brackets_on_presymplectic_manifolds}. By means of natural geometric coordinates in $T\mathbb{R}^4$, we can consider the Lagrangian function
\begin{equation}
\mathcal{L} = m \sqrt{g_{\mu \nu}\dot{x}^{\mu}\dot{x}^{\nu}}\,.
\end{equation}
For simplicity, in the following computations we set $m=1$. The associate one-form $\theta_{\mathcal{L}}$ will be 
\begin{equation}
\theta_{\mathcal{L}} = \frac{g_{\mu \nu}\dot{x}^{\mu}dx^{\nu}}{\mathcal{L}}\,.
\end{equation}
Let $v^2 = g_{\mu \nu}\dot{x}^{\mu}\dot{x}^{\nu}$, then
\begin{equation}
d\theta_{\mathcal{L}} = \omega_{\mathcal{L}} = \frac{1}{v^3} \left( g_{\mu \nu}v^2 - \dot{x}_{\mu}\dot{x}_{\nu} \right) d\dot{x}^{\nu} \wedge dx^{\mu} \,,
\end{equation} 
and we have to remove from $T\mathbb{R}^4$ the submanifold defined by $\mathcal{L}= 0$. 
We notice that $\Delta = \dot{x}^{\mu}\frac{\partial}{\partial x^{\mu}}$ and $\Gamma = \dot{x}^{\mu}\frac{\partial }{\partial x^{\mu}}$ are both in the kernel of $\omega_{\mathcal{L}}$. By passing to the quotient with respect to $\Delta$ and $\Gamma$, we get a symplectic six-dimensional manifold. To define a Poisson Bracket on functions on (the open submanifold of) $T\mathbb{R}^4$, we need to define a lift from vector fields on the quotient manifold to vector fields on $T\mathbb{R}^4$. 
At this purpose, it is possible to use a flat connection whose horizontal leaves are defined by the level sets of the functions $f_1= g_{\mu \nu}\dot{x}^{\mu} {x}^{\nu}$ and $f_2=\mathcal{L}$. A connection one-form, a $(1-1)$-tensor field $A$ which satisfies $A^2=A$ and contains $\Delta$ and $\Gamma$ in its null-space, is given by 
\begin{equation}
A = \mathbf{1} - \frac{\dot{x}_{\mu}d\dot{x}^{\mu}}{\mathcal{L}^2}\otimes \Delta - \frac{1}{\mathcal{L}} d\left( \frac{\dot{x}_{\mu}x^{\mu}}{\mathcal{L}} \right) \otimes \Gamma\,.
\end{equation}
We observe that
\begin{eqnarray}
& \frac{\dot{x}_{\mu}d\dot{x}^{\mu}}{\mathcal{L}^2}(\Delta) = 1\,,\quad \frac{1}{\mathcal{L}} d\left( \frac{\dot{x}_{\mu}x^{\mu}}{\mathcal{L}} \right) (\Delta) = 0 \\
& \frac{\dot{x}_{\mu}d\dot{x}^{\mu}}{\mathcal{L}^2}(\Gamma) = 0 \,,\quad \frac{1}{\mathcal{L}} d\left( \frac{\dot{x}_{\mu}x^{\mu}}{\mathcal{L}} \right) (\Gamma) = 1\,,
\end{eqnarray}
and that
\begin{equation}
\left[ \Delta, \Gamma \right] = \Gamma\,. 
\end{equation}
By using this connection, we define a bivector field 
\begin{equation}
\Lambda = \mathcal{L} g^{\mu \nu} \left[ A\left( \frac{\partial}{\partial \dot{x}^{\mu}} \right) \wedge A\left( \frac{\partial}{\partial x^{\nu}} \right) \right]\,,
\end{equation}
whose associated Poisson Brackets will be
\begin{eqnarray}
& \left\lbrace \dot{x}^{\rho}, {x}^{\sigma} \right\rbrace = \mathcal{L} \left( g^{\rho \sigma} - \frac{\dot{x}^{\rho}\dot{x}^{\sigma}}{\mathcal{L}^2} \right) \\
& \left\lbrace \dot{x}^{\rho}, \dot{x}^{\sigma} \right\rbrace = 0 \\
&  \left\lbrace {x}^{\rho}, {x}^{\sigma} \right\rbrace = \frac{\dot{x}^{\sigma}x^{\rho} - \dot{x}^{\rho} x^{\sigma}}{\mathcal{L}}\,.
\end{eqnarray}
Note that these brackets are very similar to those derived on the phase space with the aid of the Jacobi bracket. The  bivector field $\Lambda$ reads 
\begin{equation}
\Lambda = \mathcal{L}\left( g^{\rho \sigma} - \frac{\dot{x}^{\rho}\dot{x}^{\sigma}}{\mathcal{L}^2} \right) \frac{\partial }{\partial \dot{x}^{\rho}} \wedge \frac{\partial }{\partial x^{\sigma}}\,.
\end{equation}
This tensor field and the associated brackets are Lorentz invariant. The canonical coordinates on the symplectic quotient are given by the functions:
\begin{eqnarray}
& Q^j = \mathcal{L}\frac{K^j}{\dot{x}^0} \\
& P^j = \frac{\dot{x}^j}{\mathcal{L}}\,,
\end{eqnarray}
where we have introduced a splitting in space and time so that we may write
\begin{eqnarray}
& J_l = \frac{1}{\mathcal{L}}\epsilon_{ljk}\dot{x}^j x^k  \\
& K^j = \frac{1}{\mathcal{L}} \left( \dot{x}^j x^0 - \dot{x}^0 x^j \right)\,.
\end{eqnarray} 
We notice that the functions
\begin{equation}
Q^j = \mathcal{L}\frac{K^j}{\dot{x}^0} = -x^j + \dot{x}^j \frac{x^0}{\dot{x}^0}
\end{equation}
are the so called Newton-Wigner positions. Once more, canonical coordinates and geometrical positions do not coincide.

\begin{remark}
It is in order, at this point, to note that the $(1-1)$-tensor field 
\begin{equation}
\frac{1}{\mathcal{L}} d\left( \frac{\dot{x}_{\mu}x^{\mu}}{\mathcal{L}} \right) \otimes \Gamma
\end{equation}
behaves like a ``dynamical'' reference frame \cite{deritis_marmo_preziosi-a_new_look_at_relativity_transformations}. On the tangent bundle, however, the analog of a simultaneity submanifold would be the ``horizontal foliation'' defined by the level sets of the two functions $f_1$ and $f_2$. The dynamics takes from one simultaneity submanifold to another. If we restrict to the dynamically invariant submanifold
\begin{equation}
k_{\mu} = \frac{g_{\mu \nu}\dot{x}^{\nu}}{\mathcal{L}}\,,
\end{equation} 
we would get $f_1 = k_{\nu}x^{\nu}$ and $\frac{\Gamma}{\mathcal{L}}$ would be a reference frame in space-time in the usual meaning when $k_{\mu}k^{\mu}=1$. 

It should be noticed that, in this dynamical approach, the function
\begin{equation}
\tau = \frac{\dot{x}^{\mu}x_{\mu}}{\mathcal{L}}
\end{equation}
behaves like a dynamical time function \cite{bhamathi_sudarshan-time_as_dynamical_variable}, whereas, the dynamical vector field $\frac{\Gamma}{\mathcal{L}}$ defines ``dynamical clocks'' and $S(\Gamma)$ would give a dilation vector field along the fibers. In conclusion, the submanifold defined by the level set of the functions 
\begin{eqnarray}
& f_1 = g_{\mu \nu}\dot{x}^{\mu} {x}^{\nu} \\
& f_2 = \mathcal{L}\\
& k_{\mu} = \frac{g_{\mu \nu}\dot{x}^{\nu}}{\mathcal{L}}\,, \quad \mu = 0,1,2,3\,,
\end{eqnarray}
would give a ``dynamical simultaneity surface'' in space-time.
\end{remark}

One could consider the foliation given by the level sets of the two functions $f_1$ and $f_2$ as a ``generalized instant form'' where the instant is now given by the value of the function $f_1$ which is the dynamical time associated with the vector field $\frac{\Gamma}{\mathcal{L}}$. We close this section providing a Newtonian realization of the Poincaré algebra also in this different Lagrangian framework. Indeed, the vector fields generating boosts and rotation in $T\mathbb{R}^4$ are already tangent to the leaves of the foliation. On the other hand the generators of translations are not. However, a direct computation shows that the following vector fields solve the problem, since they are tangent to the leaves of the foliation and obey the commutation relations of the Poincaré algebra:
\begin{equation}
P_{\mu} = \frac{\partial }{\partial x^{\mu}} - \frac{x_{\mu}}{\mathcal{L}}\Gamma\,.
\end{equation}
  
The various solutions we have presented of the Dirac's problem seem to imply that, if in every description the canonical positions coincide with geometrical positions, an ``action-at-distance'' compatible with the world-line condition does not seem possible, and the intervention of the fields in the final description seems unavoidable.

\section{Conclusions}

The main idea developed in the eighties uses the constraint formalism to describe interacting multiparticle systems. One starts with a $N$-particle system by using a redundant set of variables and imposes a sufficient number of constraints (generalized mass-shell relations) in order to ensure $3N$ degrees of freedom. Several proposal were made, and a unified geometrical setting was proposed in \cite{balachandran_marmo_mukunda_nilsson_simoni_sudarshan_zaccaria-unified_geometrical_approach_to_relativistic_particle_dynamics}. 

In all these various models the description of a true physical system of $N$ particles with $3N$ degrees of freedom requires the introduction of constraints, i.e., the selection of a Poincaré invariant submanifold in the redundant initial space one has selected. In summary, the motion is generated by the constraints.

The requirement that, when a system breaks up into clusters, each of the clusters will have an evolution parameter independent of the other clusters, implies that no action-at-distance may possibly satisfy the requirement of covariance under the Poincaré group and the world-line condition\cite{balachandran_dominici_marmo_mukunda_nilsson_samuel_sudarshan_zaccaria-separability_in_relativistic_hamiltonian_particle_dynamics}. Therefore, the conclusion that can be drawn from this presentation is that a relativistically covariant description of interacting particles requires the introduction of fields. Their additional degrees of freedom, indeed, allow for the implementation of interactions which satisfy all the principles of a relativistic theory.

We close with a quotation from Sudarshan and Mukunda \cite{sudarshan_mukunda-forms_of_relativistic_dynamics_with_world_line_condition_and_separability}:
«\textit{This curious result is reminiscent of the EPR ``paradox'' in quantum theory, but the above indicated circle of ideas suggests that correlations between distant objects need not always involve transport of material influences. It may rather depend upon the indecomposable nature of the dynamical system itself. In the present context it is brought about  by the imposition of the apparently innocent WLC. As has often been shown by Dirac, there are surprising structural similarities between classical mechanics and quantum mechanics; and often ideas that were identified in quantum mechanics reappear from a deeper study of classical mechanics}».

\section*{Aknowledgements}
F.D.C. and A.I. would like to thank partial support provided by the MINECO research project MTM2017-84098-P and QUITEMAD++, S2018/TCS-A4342. A.I. and G.M. acknowledge financial support from the Spanish Ministry of Economy and Competitiveness, through the Severo Ochoa Programme for Centres of Excellence in RD(SEV-2015/0554). G.M. would like to thank the support provided by the Santander/UC3M Excellence Chair Programme 2019/2020, and he is also a member of the Gruppo Nazionale di Fisica Matematica (INDAM),Italy.

\end{document}